\documentclass[amsmath,amssymb,aps,prl,twocolumn,groupedaddress,superscriptaddress,10pt,reprint,longbibliography]{revtex4-2}

\usepackage[left]{lineno}
\usepackage{graphicx}
\usepackage{textcomp} 
\usepackage{bm}

\usepackage{exscale}

\usepackage{relsize}

\usepackage{graphicx}
\usepackage{dcolumn}
\usepackage{bm}
\usepackage[colorlinks, linkcolor=blue, citecolor=blue, urlcolor=blue, breaklinks=true]{hyperref}
\usepackage{color}
\usepackage{float}
\usepackage[
margin=0.71in,
]{geometry}

\begin{document}
\title{From local to nonlocal high-$Q$ plasmonic metasurfaces}
\author{Yao Liang}
\affiliation{Department of Electrical Engineering, City University of Hong Kong, Hong Kong, China}
\author{Din Ping Tsai}
\email{dptsai@cityu.edu.hk}
\affiliation{Department of Electrical Engineering, City University of Hong Kong, Hong Kong, China}
\affiliation{The State Key Laboratory of Terahertz and Millimeter Waves, City University of Hong Kong, Hong Kong, China}
\affiliation{Centre for Biosystems, Neuroscience and Nanotechnology, City University of Hong Kong, Hong Kong, China}
\author{Yuri Kivshar}
\email{yuri.kivshar@anu.edu.au}
\affiliation{Nonlinear Physics Center, Research School of Physics, Australian National University, Canberra ACT 2615, Australia}


\begin{abstract}
The physics of bound states in the continuum (BICs) allows to design and demonstrate optical resonant structures with large values of the quality factor ($Q$-factor) by employing dielectric structures with low losses.  However, BIC is a general wave phenomenon that should be observed in many systems, including the metal-dielectric structures supporting plasmons where the resonances are hindered by losses. Here we develop a comprehensive strategy to achieve high-$Q$ resonances in plasmonic metasurfaces by effectively tailoring the resonant modes from local and nonlocal regimes.
\end{abstract}

\maketitle

\textit{Introduction}—Recent progress in metaphotonics is driven by the physics of optical resonant modes supporting high values of the quality factor ($Q$-factor).  One of the underpinning mechanisms for high-$Q$ metaphotonics is associated with the physics of bound states in continuum (BICs) being known as spatially localized states residing in the continuum spectrum of extended states \cite{hsu2016bound}. An ideal BIC is a dark state with infinite lifetime that in practice always turns into a quasi-BIC with finite lifetime \cite{koshelev2018asymmetric}. The study of BICs and quasi-BICs has attracted a lot of attention in the recent years.  The BIC concept has been employed for many problems requiring the enhancement of light-matter interaction with numerous applications including nanolasers \cite{kodigala2017lasing,hwang2021ultralow}, high-harmonic generation \cite{zograf2022high}, biosensing \cite{leitis2019angle}, optical imaging \cite{yesilkoy2019ultrasensitive}, etc.

In a majority of applications, BICs are realized in {\it dielectric photonic structures} fabricated of materials with high value of refractive index \cite{chen2022observation,jin2019topologically,leitis2019angle,yesilkoy2019ultrasensitive}, and the underlying physics explores the idea to reduce {\it the radiation $Q$ factor} by adjusting geometric parameters such as asymmetric of meta-atoms composing metasurfaces \cite{koshelev2018asymmetric}.  At the same time, several recent studies demonstrated the use of the BIC concept for plasmonic \cite{liang2020bound,liang2021hybrid,aigner2022plasmonic} and hybrid \cite{azzam2018formation} structures.

\begin{figure}[b]
\centering
\includegraphics[width=3.375in]{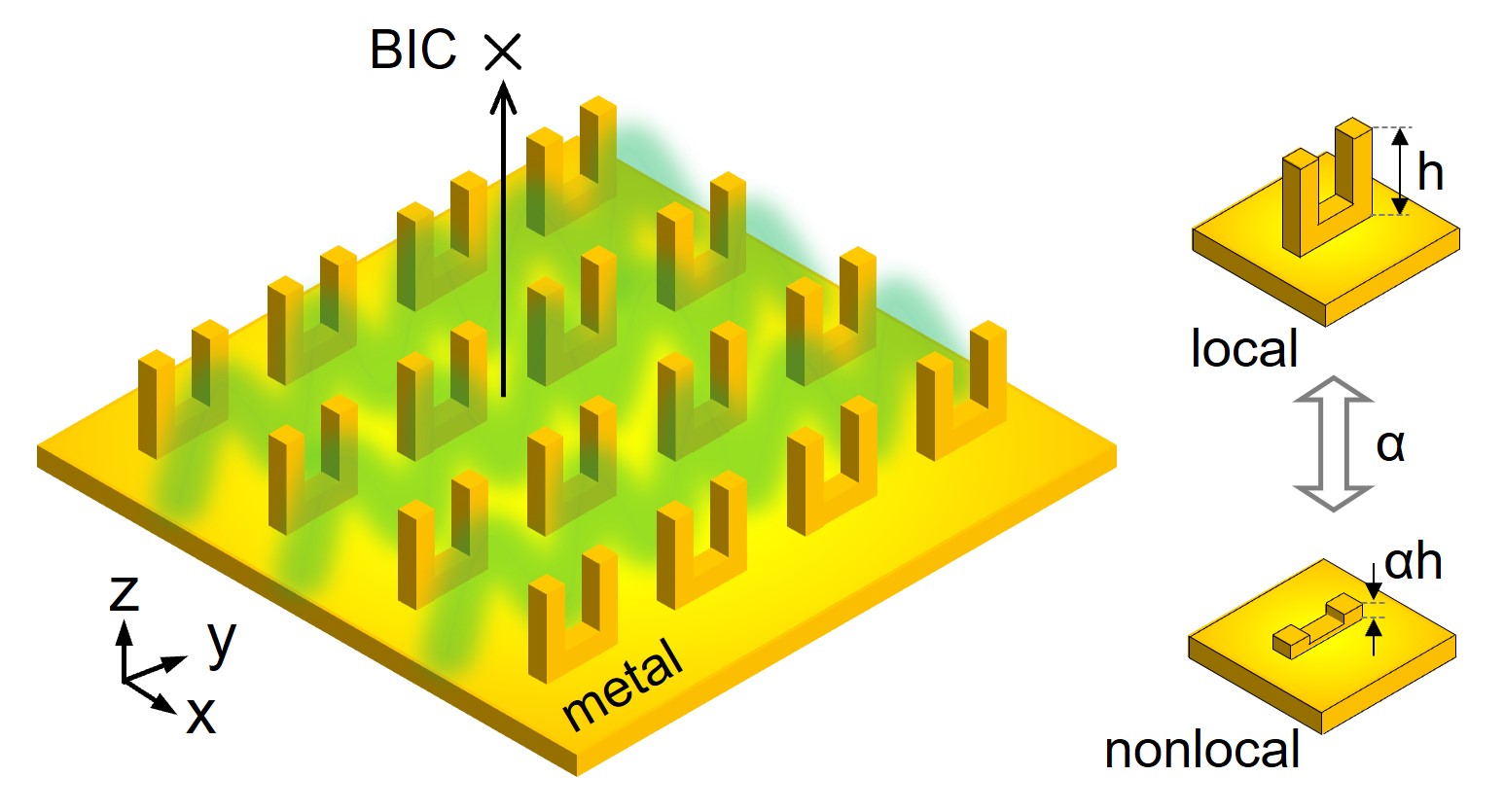}
\caption{Light-trapping in a plasmonic (gold) BIC metasurface with vertical split-ring resonators. Right:  Scaling transition between local and nonlocal resonances through the parameter scaling, with $\alpha$ being the scaling parameter.}
\label{fig_1}
\end{figure}

The BIC concept relies of the basic principles of wave physics and wave interference \cite{koshelev2023bound}; therefore, it can be applied to both low-loss dielectric and high-loss plasmonics structures. Thus, the main question is:  \textit{What the general strategy for engineering high-$Q$ resonances in plasmonic structures}?  In this Letter, we uncover the basic physics of achieving high-Q plasmonic structures via the manipulation of dissipative properties of the resonant modes during the transition between local and nonlocal regimes in plasmonic metasurfaces.

\begin{figure*}[t] 
\centering
\includegraphics[width=6in]{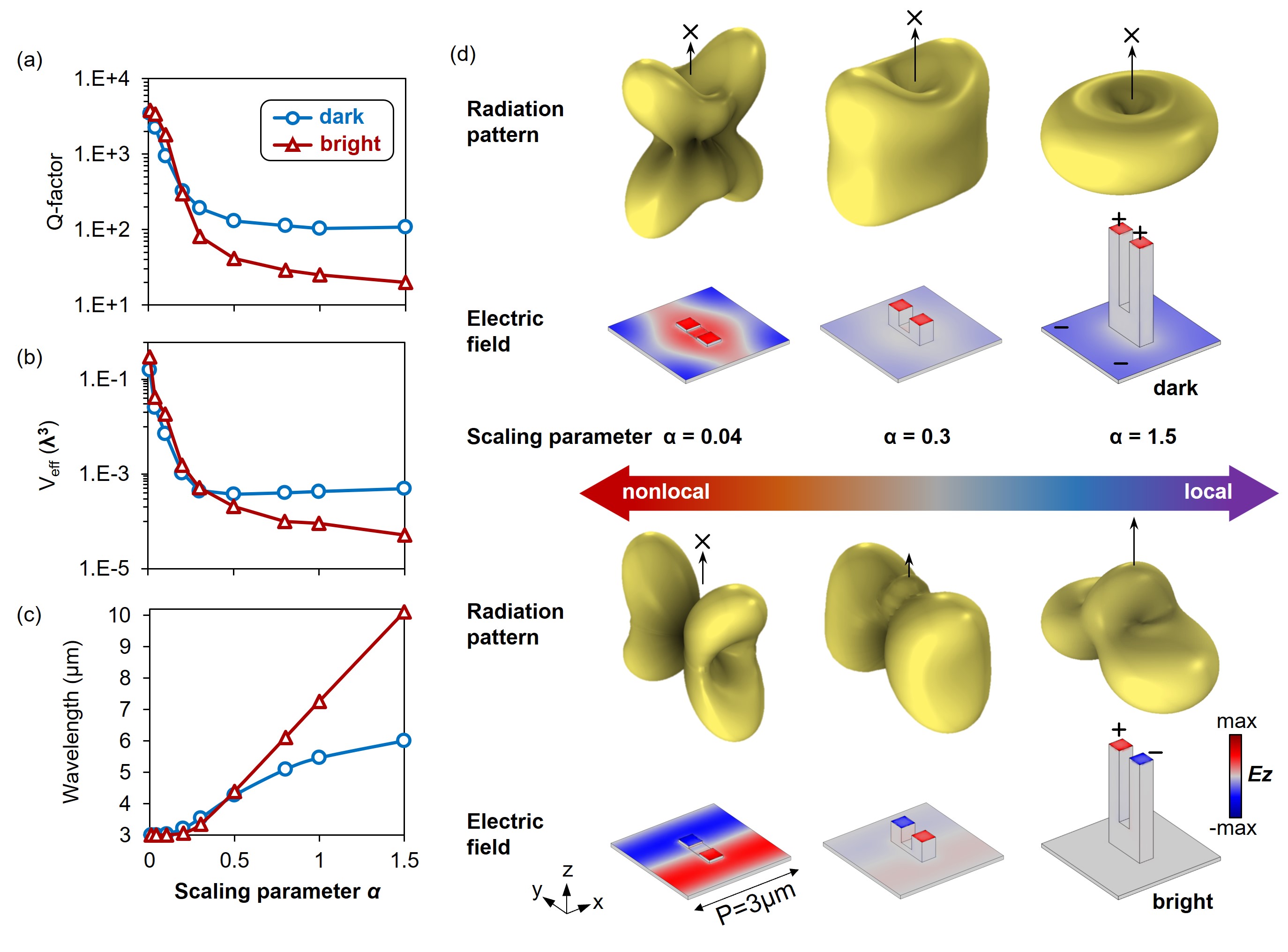}
\caption{Eigenmode analysis. (a)-(c) The $Q$-factor, mode volume, and resonance wavelength dependence on the scaling factor $\alpha$ for the dark and bright modes supported in the golden VSRR metasurface. The initial parameters (at $\alpha=1$) are period 3 $\mu m$, square pillar width 0.4 $\mu m$ and high 1.8 $\mu m$, middle connector height 0.5 $\mu m$, center-to-center distance between pillars 0.8 $\mu m$. (d) the radiation patterns and electric field (\textit{\textbf{Ez}}) distributions during the transition between local and nonlocal regimes for various scaling factors.}
\label{fig_2}
\end{figure*}

\textit{Local to nonlocal transition in the parameter space}—To illustrate our general strategy, first we focus on one recent example of a plasmonic metasurface (Fig. \ref{fig_1}), consisting of vertical split-ring resonators (VSRRs) on a golden film substrate \cite{liang2020bound}. This plasmonic structure supports dark and bright localized surface plasmon resonances (LSPRs). The initial geometric parameters are shown in the caption of Fig. \ref{fig_2}. By exclusively scaling the height parameters (the pillars and middle-connector heights) with the scaling factor $\alpha$ while keeping other parameters constant, it facilitates a transition between local and nonlocal regimes for both the modes. Here, $\alpha>1$ indicates an increase, while $\alpha<1$ signifies a decrease.\par

Figure \ref{fig_2} illustrates this transition in parameter space at ($k_x$,$k_y$)=(0,0), with $k_x$ and $k_y$ representing wave vector components along the x- and y-axes. We categorize even and odd symmetry LSPRs, corresponding to in-plane and out-of-plane resonances, as dark and bright modes based on their far-field radiation at local regimes. Decreasing the scaling parameter $\alpha$ from 1.5 to 0.01 leads to a shift from local to nonlocal resonance with 2 notable features: (1) the resonance wavelengths approach the period ($\lambda \rightarrow P$, where P = 3$\mu$m), as shown in Fig. \ref{fig_2}c; (2) A significant increase in $Q$-factor (Fig. \ref{fig_2}a), and mode volume (Fig. \ref{fig_2}b) for both modes, with differences spanning several orders of magnitude.\par

In pure local regimes, when $\alpha=1.5$, the bright LSPR mode, for example, exhibits a resonance wavelength ($\lambda\sim10\mu m$) several times larger than the lattice period ($P=3\mu m$), as shown in Fig. \ref{fig_2}c. The individual unit resonance (local) prevails in this local LSPR, overshadowing negligible contributions from collective resonances (nonlocal) that depend on strong interactions among neighboring units \cite{kravets2018plasmonic}. This is evident because a single isolated unit exhibits a nearly identical electric field profile and spectral enhancement as the entire array. (Supplemental Material (SM) \footnote{\label{SM} Supplemental Material: Simulation details, mode volume calculation, single particle resonances VS array resonances, dissipation loss control during the transition between local and nonlocal regimes, trapped SPPs, dissipation Q-factor limit of Trapped SPPs, and local and nonlocal transition for various plasmonic metasurfaces, which includes more Refs.\cite{olmon2012optical,maier2006plasmonic,krenn1999squeezing,wood1902xlii,zhou2011tunable,zijlstra2012optical,liu2010infrared,bin2021ultra,yu2010high,wood1935anomalous,joshua2021quality,sobhani2015pronounced}}, S3). \par

However, the $Q$-factor for the bright LSPR is low, $Q\approx19.8$ at $\alpha=1.5$ (Fig. \ref{fig_2}a). This is predominantly attributed to two reasons. \par

First, its tight light confinement, evident through hotspots on the tops of pillars (Fig. \ref{fig_2}d) and an ultra-small mode volume well below the diffraction limit ($V_{eff}\sim 5.13\times 10^{-5}\lambda^3$, Fig. \ref{fig_2}b). These hotspots amplify the electric field ($|\textbf{E}|$), causing a notable increase in the metal's dissipation density, $w=1/2\epsilon_0\Im (\epsilon) |\textbf{E}|^2$, where $\epsilon_0$ and $\epsilon=\epsilon_r+i\epsilon_i$ denote the vacuum permittivity and relative permittivity of gold. This giant dissipation loss hampers substaining light energy exchange between the E-field and the H-field, preventing high-Q resonances. The reason is simple: high-Q resonances, known for long-lasting light oscillation in cavities, require a sustaining oscillation between electric field energy ($u_E\propto\epsilon\textbf{E}^2$) and magnetic field energy ($u_H\propto\mu\textbf{H}^2$) in a cavity due to light's electromagnetic nature, where $\mu$ is the permeability \cite{khurgin2015deal}; Once giant E-field or H-field hotspots present, this oscillation is damaged, leading to low-Q resonances. \par

Second, it has giant radiation loss. In plasmonic cavities, their resonance Q-factor reads
\begin{equation}\label{q-factor_equation}
Q^{-1}=Q_{rad}^{-1}+Q_{dis}^{-1}
\end{equation}
where $Q_{rad}$ and $Q_{dis}$ are, respectively, the radiation and dissipation Q-factors. See SM \footnotemark[\value{footnote}]{}, S1 for simulation/calculation details.\par

One way to improve the LSPR Q-factor is to suppress radiation loss using dark modes without radiation loss, $Q_{rad}=\infty$. The Q-factor for the dark LSPR is $Q\approx107.8$ at $\alpha=1.5$ (Fig. \ref{fig_2}a), representing a 5$\times$ improvement over the bright LSPR. However, the Q-factor of the dark LSPR is still limited by the giant dissipation loss, as it has hotspots on pillars tops (Fig. \ref{fig_2}d). \par

Using our local-to-nonlocal transition strategy is an effective way to minimize dissipation loss. In this transition, the E-field becomes less confined and extends more into the lossless air (SM\footnotemark[\value{footnote}]{}, S4). This is accomplished by increasing mode volumes (Fig. $\ref{fig_2}b$) and the gradual disappearance of hotspots on pillar tops, eventually resulting in a uniformly distributed E-field profile on the gold film plane (Fig. \ref{fig_2}d). These two features substantially reduce resonances' dissipation loss, as indicated by large Q-factors, $Q\approx3439$ (dark) and $Q\approx3802$ (bright) at $\alpha=0.01$ for both modes (Fig. $\ref{fig_2}a$), several orders of magnitude larger than the local LSPRs.\par

\textit{Diffraction orders and nonlocality}—Plasmonic resonance modes in a periodic array can be decomposed into Bloch harmonics \cite{engelen2009subwavelength}, given by \(\textbf{E} (\textbf{r}) = \sum a_{(p,q)}\)$e^{-i(\textbf{G}+k_{\parallel})\textbf{r}}$, where $a_{(p,q)}$ is the complex amplitude, $k_{\parallel}=k_x+k_y$ the in-plane k-vector, $G=2\pi p/P_x + 2\pi q/P_x$ the array reciprocal vector, with the meta-unit periods $P_x=P_x=P$  and $p,q\in\mathbb{Z}$. We identify a highly symmetric position \((k_{\parallel}, \lambda)=(0,\lambda_D)\) in momentum space as the D point, where several diffraction orders degenerate, such as (p, q) = (+1, 0), (+1, 0), (0, $\pm1$), and $\lambda_D=P$ is the degenerate wavelength.\par 

Interestingly, D point ($\lambda=\lambda_D$) is a critical transition point, where the Bloch harmonics (+1, 0), (+1, 0), (0, $\pm1$) can either be stored as bounded evanescent waves when $|G|>|2\pi/ \lambda|$ ($|G|=2\pi/P$ and $\lambda>\lambda_D$), or become propagating diffraction orders when $\lambda<\lambda_D$. We involve normalized detuning wavelength ($\Delta$) to describe the distance between resonance wavelength ($\lambda$) and $\lambda_D$
\begin{equation}
    \Delta=\frac{\lambda-\lambda_D}{\lambda_D}
    \label{delta}
\end{equation}\par
As $\Delta$ decreases and approaches 0, the LSPRs (strong light confinement) gradually turn into surface plasmon polaritons (SPPs) with E-field enormously extending into the air due to their strong coupling with the nonlocal diffraction orders. This explains why the nonlocal modes have mode volumes orders of magnitude bigger than the local LSPRs (Fig. \ref{fig_2}). 

\begin{figure}[t]
\centering
\includegraphics[width=3.375in]{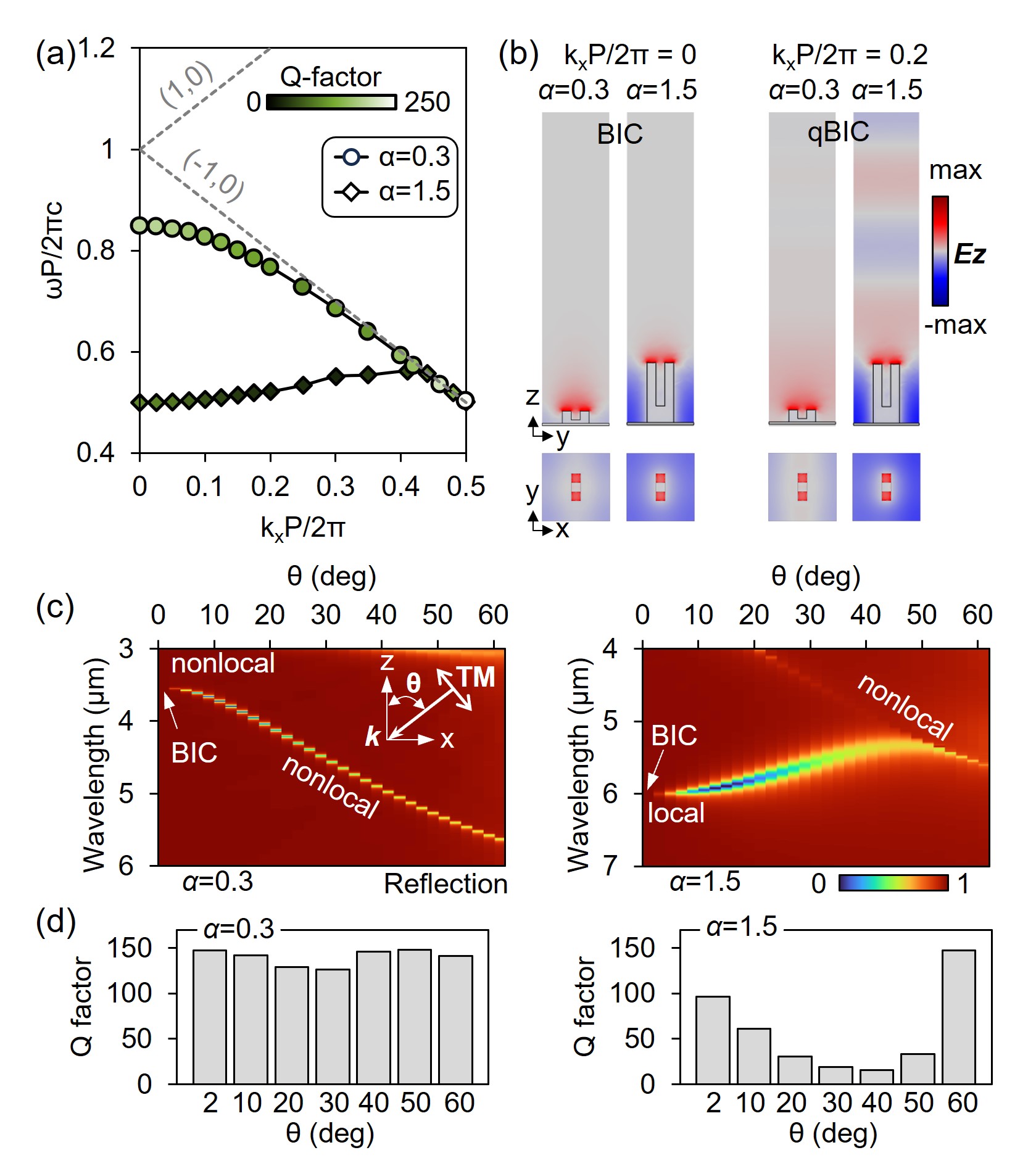}
\caption{(a) Calculated band structure for the dark mode with nonlocal ($\alpha=0.3$) and local ($\alpha=1.5$) characteristics, denoted by circular and square markers, respectively. The color represents  eigenmode Q-factor. The gray dashed lines represent two diffraction orders, (+1, 0) and (-1, 0). (b) The field distribution of dark modes for two metasurfaces ($\alpha=0.3$ and $\alpha=1.5$) at $k_xP/2\pi=0$ ($\Gamma$ point) and $k_xP/2\pi=0.2$ (off-$\Gamma$). (c) Reflection spectra for two metasurfaces ($\alpha=0.3$ and $\alpha=1.5$) at oblique incidence in the x-z plane. (d) The corresponding Q-factor of the two dark modes at different oblique angles.}
\label{fig_3}
\end{figure}

\textit{Nonlocal modes nature}—The high-$Q$ nonlocal modes are trapped SPPs, characterized by standing SPP waves confined in a Fabry–P\'erot cavity (SM \footnotemark[\value{footnote}]{}, S4, S5). Two pieces of evidence support this interpretation. \par

First, trapped SPPs exhibit no far-field radiation. Consequently, all nonlocal plasmonic modes, whether transitioning from a dark or bright LSPR in the local to nonlocal shift, should remain sub-radiative if they are trapped SPPs. The dark LSPR supports this characteristic throughout the transition (Fig. \ref{fig_2}d). Interestingly, despite being radiative in local regimes ($\alpha=1.5$), the bright LSPR becomes less radiative as $\alpha$ decreases to 0.3 and eventually becomes radiation-free in nonlocal regimes ($\alpha\leq0.04$) (see Fig. \ref{fig_2}d and S3 in SM\footnotemark[\value{footnote}]{}). This alignment with the dark feature of trapped SPPs. Second, another evidence is linked to the Q-factor limit of the nonlocal mode.\par

\textit{Q-factor limit}—As $\Delta$ decreases, dark and bright LSPRs shift into trapped SPPs, exhibiting minimal dissipation loss, enabling efficient energy exchange between E-field and H-field. To determine the upper limit of Q-factors for nonlocal plasmonic resonances we can assess:
\begin{equation}
    Q_{max}=k_{spp}^{r}/2k_{spp}^{i}
    \label{q-max}
\end{equation}
where $k_{spp}^{r}$ and $k_{spp}^{i}$ are the real and imaginary part of SPPs k-vector, such that $k_{spp}=k_{spp}^{r}+ik_{spp}^{i}=\frac{2\pi}{\lambda}\sqrt{\frac{\epsilon\epsilon_0}{\epsilon+\epsilon_0}}$, where $\epsilon$ and $\epsilon_0$ are the permittivities of gold and vacuum \cite{zayats2005nano}. As $\epsilon$ varies with wavelength, the maximum $Q$-factor of nonlocal plasmonic resonance is wavelength-dependent (SM \footnotemark[\value{footnote}]{}, S5). For $\lambda\approx3\mu$m, the trapped SPPs Q-factor is calculated as $\sim3805$, consistent with numerical results of nonlocal mode in Fig. ~\ref{fig_2}a.\par
 
\textit{Local-nonlocal transition in the momentum space}— The local-to-nonlocal transition can also occur in the momentum space as the resonance modes interact with nonlocal diffraction orders. For example, we calculate the eigenfrequency and Q-factor for two dark modes with different height parameters, $\alpha=1.5$ (local) and $\alpha=0.3$ (nonlocal), as shown in Fig. \ref{fig_3}a. They both symmetry-protected BICs with zero radiation loss at the $\Gamma$-point ($k_{\parallel}=0$), as shown in Fig. \ref{fig_2}b (left). As the in-plane vector $k_x$ increases, BICs transit to quasi-BICs with several distinguished properties: \par

First, the Q-factor of the local quasi-BIC mode experiences a rapid decrease, whereas the Q-factor of the nonlocal quasi-BIC mode remains stable, as verified by both eigenmode studies (Fig. \ref{fig_3}a) and full-wave simulations (Figs. \ref{fig_3}c,d). This can be explained in terms of coupling strength between the resonance modes and nonlocal diffraction order (-1, 0). For example, the local mode is closer to the (-1, 0) with a smaller $\Delta$ than its nonlocal counterpart at $k_xP/2\pi=0.2$. Thus, it relies more on the mutual interaction among neighboring units, which reduces its radiation loss \cite{kravets2018plasmonic}. This makes it less radiative compared to the local counterpart (Fig. \ref{fig_3}b, right). Also, its nonlocal feature makes it less dissipative. These two features help it keep high-Q resonances at various oblique incidences (Fig. \ref{fig_3}c,d). \par

Second, the local mode becomes nonlocal at large oblique incidence angles ($\theta > 40^{\circ}$) as the detuning wavelength decreases. This is evident by a sudden increase in its Q-factor when $\theta > 40^{\circ}$ (Fig. \ref{fig_3}d).\par

\begin{figure}[t]
\centering
\includegraphics[width=3.375in]{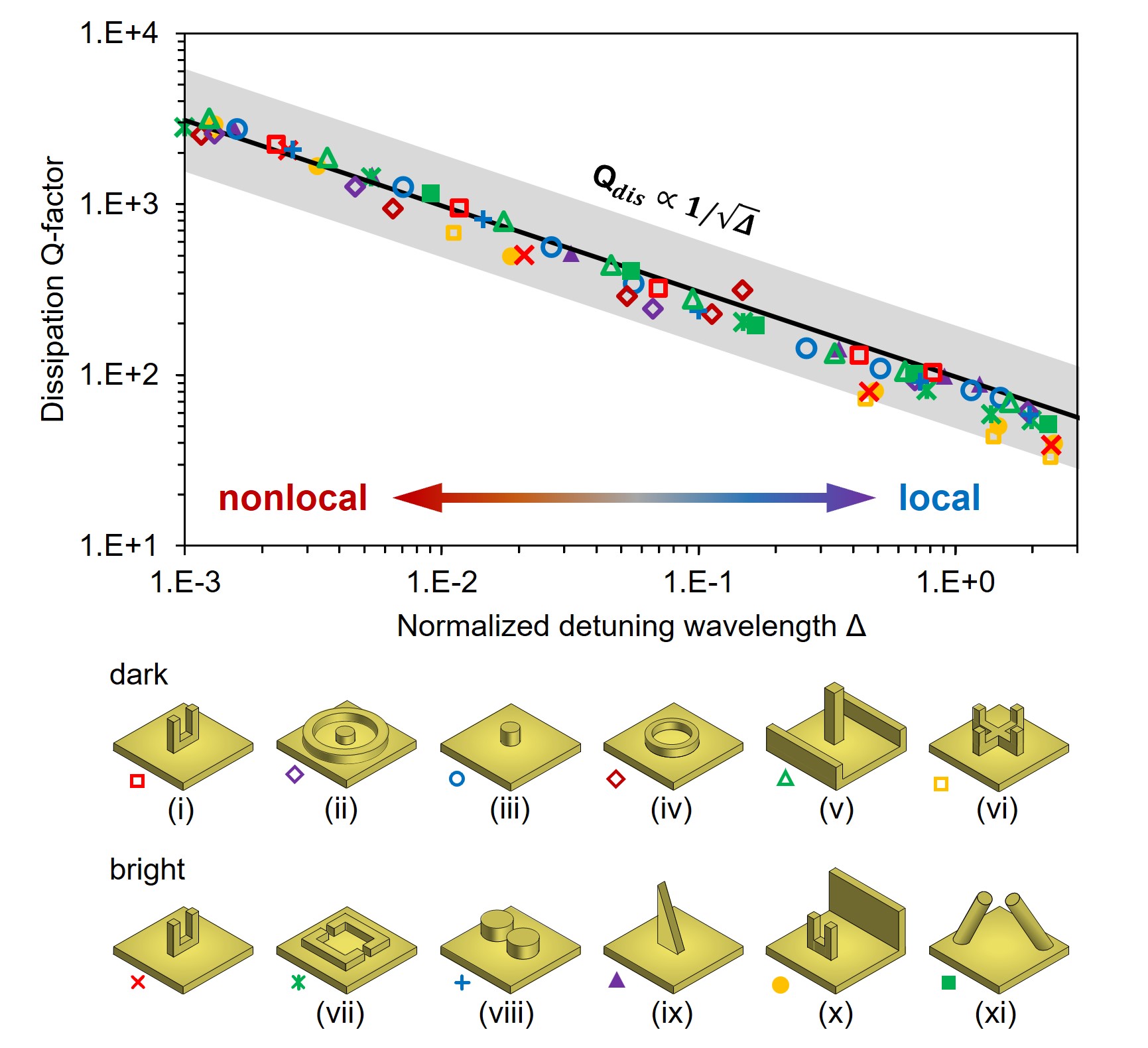}
\caption{Dependence of the dissipation $Q$ factor on the normalized detuning wavelength $\Delta$ (log-log scale) for various all-plasmonic designs [(i) to (xi)] supports bright/dark LSPRs (Refs. \cite{liang2020bound,yang2015unidirectional,aigner2022plasmonic,cetin2012fano,gerislioglu2020monolithic,li2014ultranarrow,liang2021hybrid,shen2022terahertz,tang2023chiral,xiong2013structured,zilio2015hybridization}) All square units have a 3$\mu$m period.}
\label{fig_4}
\end{figure}

\textit{Local-nonlocal transition in plasmonic metasurfaces}—Our approach to enhancing the Q-factor in plasmonic nanostructures by reducing the height parameter is not limited to a specific metasurface with VSRR units. Instead, it is a universally applicable strategy that can be employed for all types of plasmonic metasurfaces with various meta-units, including single pillar, ring, dimmer, verticle triangle, pillar-wall, and many others \cite{liang2020bound,yang2015unidirectional,aigner2022plasmonic,cetin2012fano,gerislioglu2020monolithic,li2014ultranarrow,liang2021hybrid,shen2022terahertz,tang2023chiral,xiong2013structured,zilio2015hybridization}, as shown in Fig. \ref{fig_4}. \par
All meta-atoms in Fig. \ref{fig_4} can support LSPRs. For simplicity in our discussion, we set them square units with a period of 3$\mu m$, and their initial height as 1.8 $\mu m$ (at $\alpha=1$). Thus, their degenerate wavelength, where (+1, 0), (+1, 0), (0, $\pm1$) diffraction orders emerge, is $\lambda_D =3 \mu m$. As the height scaling parameter $\alpha$ decreases, the LSPRs (dark or bright) begin to transition into trapped SPPs with similar E-field profiles, whether exhibiting symmetry or anti-symmetry (SM \footnotemark[\value{footnote}]{}, S7).\par
We study the dependence of the dissipation Q-factor $Q_{dis}$ on the normalized detuning wavelength $\Delta$ for all plasmonic metasurfaces during the local (LSPRs) to nonlocal (trapped SPPs) transition (Fig. \ref{fig_4}). An inverse square root law well approximates the relationship
\begin{equation}
    Q_{dis} \propto \frac{1}{\sqrt{\Delta}}
    \label{q-dis}
\end{equation}
where $\Delta$ is calculated using Eq. \ref{delta}, and the dissipation Q-factor using Eq. \ref{q-factor_equation} (SM \footnotemark[\value{footnote}]{}, S1). Notably, in the nonlocal regimes ($\alpha \sim 10^{-3}$), the Q-factor ($Q=Q_{dis}$) of all plasmonic metasurfaces is approaching $\sim3800$, consistent with prediction using Eq. \ref{q-max}, confirming the trapped SPPs nature of the nonlocal modes.\par
Although drawing a clear line between local and nonlocal regimes is difficult, the inverse square root law suggests an intelligent way to engineer plasmonic structures with on-demand resonances during this transition. For example, most resonances have hybrid (LSPRs + SPPs) properties during the local-to-nonlocal transition. For example, at $\alpha=0.3$, the bright mode has both hotspots (local) on pillars tops and trapped SPPs on the ground plane (Fig. \ref{fig_2}d). Notably, hotspots \cite{tsai1994photon} and high-Q resonances \cite{liang2020bound} are effective ways to enhance the electromagnetic field. Consequently, the maximum E-field intensity occurs at some point during this transition (SM \footnotemark[\value{footnote}]{}, S3), which proves beneficial for applications such as nonlinear enhancement \cite{czaplicki2018less}. \par

Equation~(\ref{q-dis}) is valid for a broad range, $\Delta \in [10^{-3}, 1]$, allowing diverse Q-factor choices for most plasmonic metasurfaces, ranging from tens to thousands. However, the plasmonic Q-factor has inherent limits, approaching that of trapped SPPs (Eq. \ref{q-max}). To attain higher Q-factors, selecting a longer operational wavelength is crucial, as the maximum Q-factor is wavelength-dependent. This is exemplified by $Q_{max}\sim 627$ at $\lambda=880nm$ and $Q_{max}\sim 6330$ at $\lambda=5\mu m$ (SM \footnotemark[\value{footnote}]{}, S6). This wavelength-dependent trend aligns with recent experimental findings ($\sim80$ in near-IR \cite{yang2015unidirectional} and $\sim500$ in mid-IR \cite{li2014ultra}) \par

\textit{Conclusion}—We have suggested and demonstrated a general conceptual approach for achieving large $Q$ factors in plasmonic metasurfaces by engineering dissipative losses and dissipation $Q$ factor of the resonant modes. Our approach employs plasmonic local and extended resonances based on the BIC physics, and it is underpinned by an efficient control of local and nonlocal optical response.  We believe our approach opens the door to many applications of high-$Q$ plasmonic structures including subwavelength lasers, harmonic generation, biosensing, optical imaging, entangled photon generation, etc. \par

\begin{acknowledgments}
This work is supported by the University Grants Committee / Research Grants Council of the Hong Kong Special Administrative Region, China [Project No. AoE/P-502/20, CRF Project: C1015-21E; C5031-22G; and GRF Project: CityU15303521; CityU11305223; CityU11310522; CityU11300123], the Department of Science and Technology of Guangdong Province [Project No. 2020B1515120073], City University of Hong Kong [Project No. 9380131, 9610628, and 7005867], and the Australian Research Council (grant DP210101292).
\end{acknowledgments}


\bibliography{maintext}

\end{document}